\documentclass{PoS}

\title{Diffractive open charm production from the dipole model analysis}

\ShortTitle{Diffractive charm production}

\author{\speaker{Agnieszka \L{}uszczak}\thanks{This work is partially supported by the grant of MNiSW 
N202 249235.}\\
        Institute of Nuclear Physics Polish Academy of Sciences, Cracow, Poland\\
        E-mail: \email{agnieszka.luszczak@ifj.edu.pl}}

\abstract{We present a precise comparison of the results of the dipole models
with the newest data from HERA on the diffractive open charm production. We found good agreement with the data on the diffractive open charm production both for the the gluon distributions from the considered dipole models and the DGLAP fits to HERA data from our earlier analysis for diffractive parton distributions with higher twist. We also show that exclusive diffractive charm production can practically be neglected.}

\FullConference{European Physical Society Europhysics Conference on High Energy Physics,
EPS-HEP 2009,\\
		 July 16 - 22 2009\\
		 Krakow, Poland}

\begin{document}

\newcommand{\beeq}{\begin{eqnarray}}
\newcommand{\eeeq}{\end{eqnarray}}
\newcommand{\be}{\begin{equation}}
\newcommand{\ee}{\end{equation}}
\newcommand{\bea}{\begin{array}}
\newcommand{\eea}{\end{array}}

\newcommand{\eq}{&=&}
\newcommand{\eto}{{\mbox{\textrm e}}}

\def\xp{x_{{I\!\!P}}}
\def\alfas{{\alpha_s}}
\def\qbar{\overline{q}}
\def\cbar{\overline{c}}
\def\bbar{\overline{b}}
\def\sigmahat{\hat{\sigma}}
\def\Qbar{\overline{Q}}
\def\half{\textstyle{\frac{1}{2}}}
\def\gev{{\textrm GeV}}
\def\rbo{{\bf r}}
\def\bbo{{\bf b}}

\section{Introduction}
In this analysis, we consider two  important parameterisations of the dipole scattering amplitude, 
called GBW  \cite{Golec-Biernat:1998js} and CGC  \cite{Soyez:2007kg}, in which parton saturation results are built in.
The comparison we performed prompt us to discuss some subtle points of the dipole
models, mostly related to the $q\qbar g$ component, and connect them to the approach based
on the diffractive parton distributions evolved with the Dokshitzer-Gribov-Lipatov-Altarelli-Parisi (DGLAP) equations. Within the latter approach,  the  diffractive open charm production is particularly interesting
since it is sensitive to a diffractive gluon distribution. However, the accuracy of the existing
data on such a production does not allow to discriminate between different gluon distributions considered in
our analysis.

\section{Diffractive charm production in dipole model}

\begin{figure*}[t]
\centerline{
         \includegraphics[width=6.5cm]{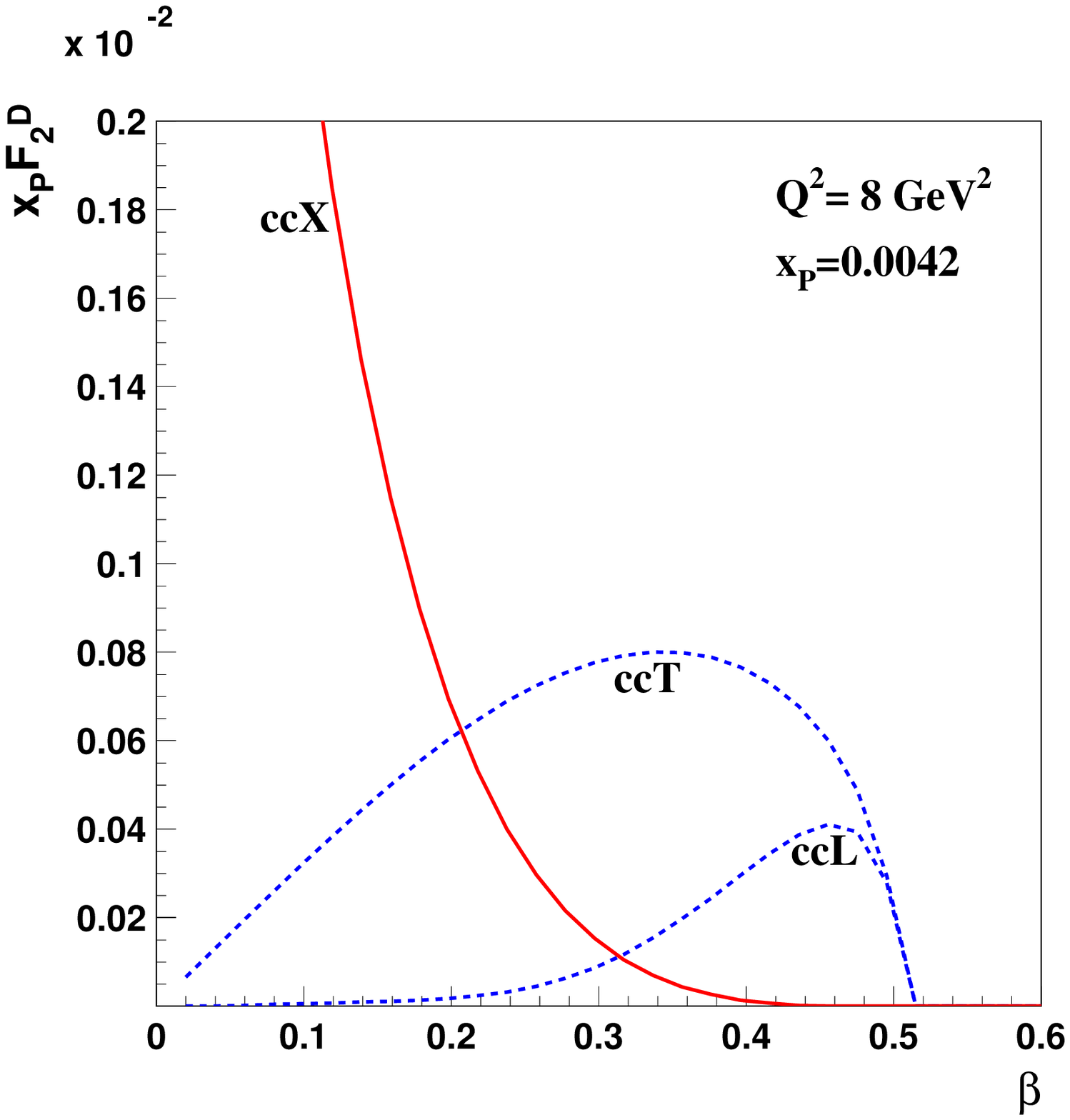}
         \includegraphics[width=6.5cm]{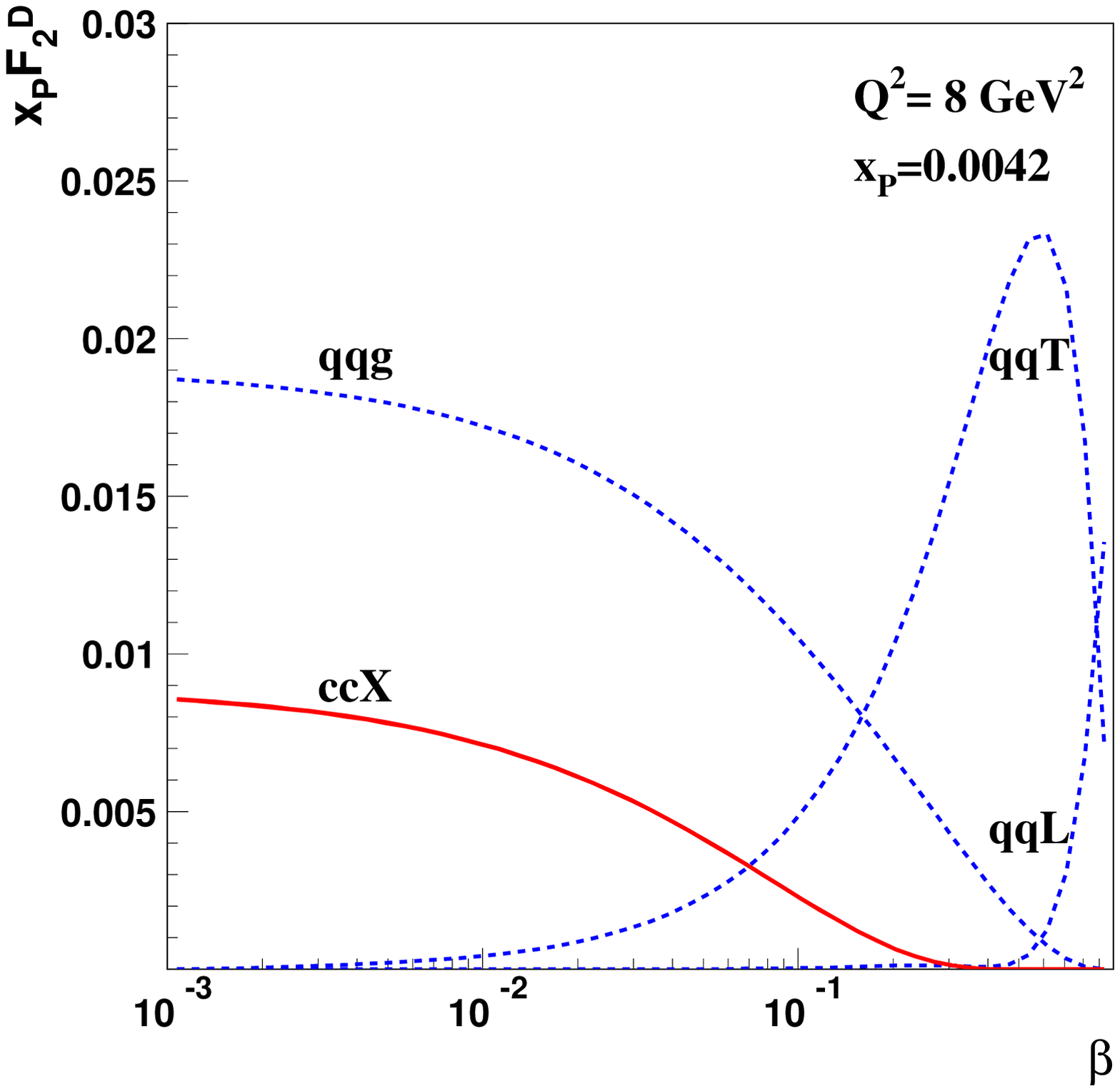}
            }
\caption{Tree components of $F_2^D$ from the dipole model with GBW parametrization.}
\label{fig:4}
\end{figure*}

In the diffractive scattering heavy quarks are produced 
in quark-antiquark pairs, $c\cbar$ and $b\bbar$ for charm and bottom, respectively. 
Such pairs can be produced provided that 
the diffractive mass of is above the quark pair production threshold
\be\label{eq:12}
M^2=Q^2\left(\frac{1}{\beta}-1\right) > 4m_{c,b}^2
\ee
In the lowest order the diffractive state consist only the  $c\cbar$ or $b\bbar$ pair.
In the forthcoming we consider only charm production since bottom production is negligible.
For example, for charm production from transverse photons we have 
\beeq\nonumber
\label{eq:5new}
\xp F_T^{(c\cbar)}\!\!\eq\!\!
\frac{3 Q^4 e_c^2}{64\/\pi^4\beta B_d} \int_{z_{c}}^{1/2}
dz\, z(1-z)
\\
&\times&\left\{
[z^2+(1-z)^2]\,Q^2_c\,\phi_1^2 \,+\,m_c^2\, \phi_0^2
\right\}
\eeeq
where $m_c$ and $e_c$ are charm quark mass and electric charge, respectively.
The minimal value of diffractive mass equals:
$M^2_{min}=4m_c^2$, thus the maximal value of $\beta$ is given by
\be
\beta_{max}=\frac{Q^2}{Q^2+4m_c^2}\,.
\ee
In such a case,  $z_c=1/2$ in eq.~(\ref{eq:5new})
and  $F_{T,L}^{(c\cbar)}=0$ for $\beta> \beta_{max}$. This is shown in Fig.~\ref{fig:4} where on the left side we have: the ${c\cbar} T$ and $c\cbar L$ components of $F_2^D$ from the dipole model with the GBW parameterisation together with the $c\cbar X$ contribution from the collinear factorisation approach (\ref{eq:13}) with the  diffractive gluon distribution \cite{GolecBiernat:2007kv} and respectively on the right: the $c\cbar X$ component in a different scale against the massless $q\qbar T$, $q\qbar L$ and $q\qbar g$ components. By the comparison with the corresponding curves 
for three massless quarks $(q\qbar T,q\qbar L, q\qbar g)$, shown in Fig.~\ref{fig:4} (right), we see that the 
exclusive diffractive charm production  contributes only $1/30$ to the total structure function $F_2^D$. Thus it can practically be neglected.

The next component is the $c\cbar g$ 
diffractive state. Unfortunately, formula for the $q\qbar g$ 
production is only known in the massless quark case and  cannot be used for heavy quarks.
Thus, we have to resort to the collinear factorisation formula, given by eq.~(\ref{eq:13}),
in which the charm-anticharm  pair
is produced via the photon-gluon fusion: $\gamma^*g\to c\cbar$ \cite{Goncalves:2004dv}. If such an approach is
applied to diffractive scattering,  gluon is a ``constituent of a pomeron''. The diffractive state consists 
of additional particles $X$ (called ``pomeron remnant'') 
in addition to the heavy quark pair,  which are well separated in rapidity from the scattered  proton.
The collinear factorisation formula for the charm contribution to the diffractive structure functions is 
taken from the fully inclusive case \cite{Gluck:1994uf} in which the standard gluon distribution is replaced by the diffractive gluon distribution $g^D$:
\beeq
\nonumber
\label{eq:13}
\xp F_{2,L}^{D(c\overline{c}X)}\!\!\eq\!\!
2\beta\, e_c^2\, \frac{\alpha_s(\mu_c^2)}{2 \pi}
\int_{a\beta}^1\frac{dz}{z}\,
C_{2,L}\!\left( \frac{\beta}{z},{\frac{m_c^2}{Q^2}} \right) 
\\
&\times&\xp g^D(\xp,z,\mu_c^2)
\eeeq
where $a = 1 + 4 m_c^2/Q^2$ and the factorisation scale $\mu_c^2=4 m_c^2$
with the charm quark mass  $m_c=1.4~\gev$.  The leading order coefficient functions are given by
\beeq\nonumber
C_2(z, r)\!\! \eq\!\!  \half\left\{z^2 + (1-z)^2 + 4z(1-3z) r -
8z^2 r^2\right\} 
\\\nonumber
&\times& \ln{\frac{1+\alpha}{1 -\alpha}} 
\,+\,\half \alpha\big\{ -1 + 8z(1-z)
\\
&-& 4z(1-z)r \big\}
\\ \nonumber
\\\label{eq:14b}
C_L(z, r)\!\!  \eq\!\!  - 4z^2 r \ln{{1+\alpha}{1 -\alpha}} + 2\alpha z (1 - z)
\eeeq
where $r={m_c^2}/{Q^2}$ and $\alpha=\sqrt{1 -{4 r z}/{(1-z)}}$.
The lower integration limit in eq.~(\ref{eq:13}) results from the condition
for the heavy quark production in the fusion:  $\gamma^*g\to c\cbar$,
\be
(z\xp\, p +q)^2\ge 4m_c^2
\ee
where we assume that gluon carries a fraction $z$ of the pomeron momentum $\xp p$.

\begin{figure*}[t]
\begin{center}
\includegraphics[width=13cm]{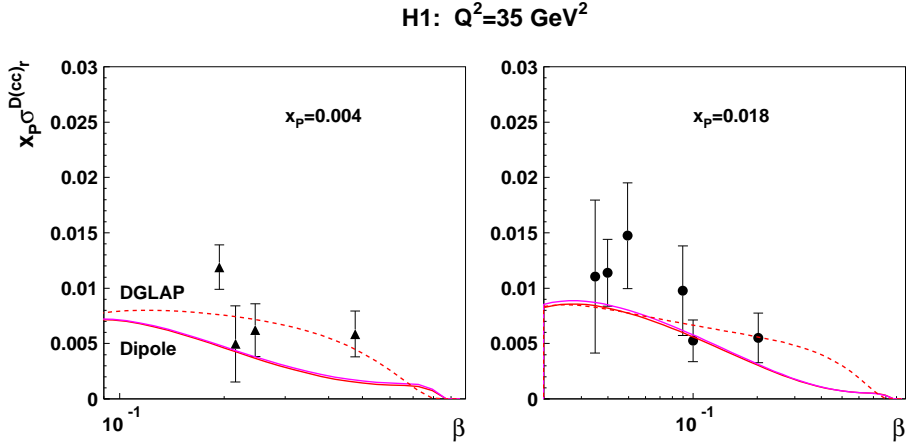}
\caption{A comparison of the collinear factorisation predictions with the GBW and
CGC gluon distributions (solid lines) with the HERA data on the open diffractive charm production.
The dashed lines are computed with the  gluon distribution obtained in the  DGLAP fit \cite{GolecBiernat:2007kv} to the H1 data on the diffractive structure functions.}
\label{fig:6}
\end{center}
\end{figure*}

The $c\cbar X$ contribution given by eq.~(\ref{eq:13}) is shown in  Fig.~\ref{fig:4} as the solid lines. 
As seen in the left figure, this component becomes significant for $\beta<0.1$ . By a comparison with the massless quark contributions (the right figure) we see that  diffractive charm production contributes up to $30\%$ to the diffractive structure function $F_2^D$ for small values of $\beta$. The presented results were obtained assuming the diffractive
gluon distribution which results from the dipole models, given in our last paper \cite{GolecBiernat:2008gk}, with the GBW
parameterisation of the dipole cross section and with the color factor modification. The CGC parameterisation gives a similar result.

In Fig.~\ref{fig:6} we show the collinear factorisation predictions for the diffractive charm production confronted with the new HERA data \cite{Aktas:2006up} on the charm component of the reduced cross section:
\be
\sigma_r^{D(c\cbar)}=F_2^{D(c\cbar)}-\frac{y^2}{1+(1-y)^2}F_L^{D(c\cbar)}\,.
\ee
The solid curves, which are barley distinguishable, correspond to the 
result with the GBW and CGC parameterisations of the diffractive gluon distributions.
The dashed lines are computed for the gluon distribution from a fit  to the H1 data \cite{GolecBiernat:2007kv} based on the DGLAP equations. 
The present accuracy of the charm data does not allow to discriminate between these two approaches although
the data seem to prefer the gluon distribution from the DGLAP fit which is much more
concentrated in the large $z$-region as compared to the dipole model gluon distributions,
see \cite{GolecBiernat:2008gk} in Appendix.

\begin{figure*}[t]
\begin{center}
\includegraphics[width=12cm]{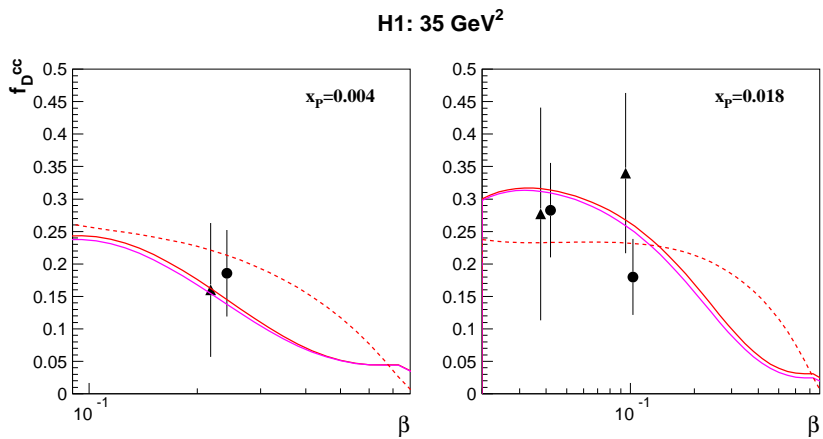}
\caption{The fractional charm contribution}
\label{fig:7}
\end{center}
\end{figure*}

The importance of diffractive charm is illustrated in Fig.~\ref{fig:7} where the fractional charm contribution,
\be\label{eq:fraccc}
f_D^{c\cbar}={\sigma_r^{D(c\cbar)}}/{\sigma_r^{D}}\,,
\ee
to the total diffractive cross section, is shown as a function of $\beta$ for two values of $\xp=0.004$ and $0.018$ against the H1 collaboration data \cite{Aktas:2006up}.The solid lines
are computed for  the $c\cbar X$ contribution  with the GBW and CGC diffractive gluon distributions while the dashed lines are found for the  diffractive gluon distribution obtained in the  DGLAP fit \cite{GolecBiernat:2007kv} to the H1 collaboration data.
For small values of $\beta$, the charm contribution equals on average approximately
$20-30\%$, which is comparable to the charm fraction in the inclusive cross section  for similar values of $Q^2$ \cite{Aktas:2005iw}. 
\section{Acknowledgments}
I am indebted to Krzysztof Golec-Biernat for collaboration on the subject of this presentation.


\end{document}